\documentclass[showpacs,superscriptaddress,twocolumn,amsmath,aps]{revtex4}
\usepackage{graphicx,color}
\usepackage{bm}
\usepackage[hypertex]{hyperref}

\newcommand{\be}{\begin{equation}}
\newcommand{\ee}{\end{equation}}
\newcommand{\bea}{\begin{eqnarray}}
\newcommand{\eea}{\end{eqnarray}}
\newcommand{\bsube}{\begin{subequations}}
\newcommand{\esube}{\end{subequations}}

\newcommand{\Eq}[1]{Eq.\,(\ref{#1})}
\newcommand{\Eqs}[1]{Eqs.\,(\ref{#1})}

\newcommand{\dg}{\dagger}

\usepackage{amsfonts}
\usepackage{amsmath}
\usepackage{amssymb}
\usepackage{graphicx}
\usepackage{type1cm}        
\usepackage{times}          
\usepackage{bm}
\usepackage{CJK}            
\usepackage{color}                      

\newcommand{\bet}{\beta}
\newcommand{\gam}{\gamma}
\newcommand{\del}{\delta}
\newcommand{\eps}{\epsilon}

\newcommand{\ome}{\omega}
\newcommand{\Gam}{\Gamma}
\newcommand{\Del}{\Delta}

\newcommand{\Lam}{\Lambda}

\newcommand{\Ome}{\Omega}

\newcommand{\beq}{\begin{equation}}
\newcommand{\eeq}{\end{equation}}
\newcommand{\beqn}{\begin{eqnarray}}
\newcommand{\eeqn}{\end{eqnarray}}
\newcommand{\bsub}{\begin{subequations}}
\newcommand{\esub}{\end{subequations}}
\newcommand{\iny}{{\infty}}
\newcommand{\re}{\nonumber\\}
\newcommand{\ket}[1]{{\left| #1 \right\rangle }}

\newcommand{\adg}{a^\dagger}

\begin{document}
\begin{CJK*}{GBK}{song}

\title{Non-Markovian Transmission through Two Quantum Dots Connected by a Continuum}

\author{Yunshan Cao}
\affiliation{School of Physics, Peking University, Beijing 100871,China}
\author{Luting Xu}
\affiliation{Department of Physics, Beijing Normal University,
Beijing 100875, China}
\author{Jianyu Meng}
\affiliation{Department of Physics, Beijing Normal University,
Beijing 100875, China}
\author{Xin-Qi Li}
\email{lixinqi@bnu.edu.cn}
\affiliation{Department of Physics, Beijing Normal University,
Beijing 100875, China}

\date{\today}

\begin{abstract}
We consider a transport setup
containing a double-dot connected by a continuum.
Via an exact solution of the time-dependent Schr\"odinger equation,
we demonstrate a highly non-Markovian {\it quantum-coherence-mediated}
transport through this dot-continuum-dot (DCD) system,
which is in contrast with the common premise since in typical case
a quantum particle does not reenter the system of interest
once it irreversibly decayed
into a continuum (such as the spontaneous emission of a photon).
We also find that this DCD system supports an unusual steady state
with unequal source and drain currents, owing to electrons
irreversibly entering the continuum and floating there.
\end{abstract}

\pacs{03.65.Yz,42.50.-p,73.23.-b}

\maketitle



Coupling of a finite-size {\it system} to a {\it reservoir}
with continuum energy spectrum is relevant to
many fundamental issues of physics,
e.g., the emergence of classicality from a quantum world \cite{Zur09},
and the origin of {\it irreversibility} in statistical physics.
Simple example of such type of coupling can be a small system,
with discrete states,
coupled to a {\it continuum} in terms of quantum tunneling.

In most cases, coupling with a continuum would result in
an {\it irreversible} decay of the discrete state,
with the particle never coming back,
such as the spontaneous emission of photon from an atom.
However, in a recent study \cite{SG11a}, where a double-dot
connected by a continuum was considered, it was found that
the second dot can affect the decay from the first
dot via quantum interference, and finally lead to
a formation of stationary bound state embedded in the {\it continuum}.
In subsequent related studies \cite{SG11b,SG11c},
also concerning this dot-continuum-dot (DCD) system,
it was demonstrated that the electron's motion through the continuum
is very unusual, say, revealing an undetectable feature
in the transfer process, and the information lost into the continuum
can be retrieved without generating more disturbance.

In the present work we consider inserting the DCD system into
a transport configuration, as schematically shown by Fig.\ 1,
which allows for an investigation for the behavior of
{\it continuous current} through the DCD setup.
Moreover, beyond the above DCD studies \cite{SG11a,SG11b,SG11c},
where a wide-band-limit (WBL) model
was assumed for the central continuum, we will generalize our study to a
finite-bandwidth Lorentzian spectrum (FBLS) for both the leads and the continuum.
As a consequence, the {\it transient} transport process is
highly {\it non-Markovian}, much stronger than in the WBL \cite{note-1}.
Particularly, associated with the peculiar DCD setup,
we will discuss and uncover other features
in relation with the {\it coherence-mediated} transfer
and electron accumulation (floating) in the continuum,
basing both on an {\it exact} treatment for the non-Markovian transport.
The issue of non-Markovian transport through quantum dots
has received considerable attention in the past years
\cite{Kon06,Jau08,Jau10,Bran04,Bran11,Bran09,Bran11a,Jin11},
where the non-Markovian transient dynamics are usually manifested
in the current noise spectrum or in terms of full counting statistics.
Moreover, exploiting the nature of non-Markovianity in broader contexts
has been a subject of great interest in recent years,
which stimulated novel means such as information-theoretic approach
to recognize that non-Markovian dynamics is a reversed flow of {\it information}
from the environment back to the open system \cite{Br09,Cir08}.
Also, experimental progress on revealing the non-Markovian effects
was reported very recently \cite{Guo11}.

\begin{figure}[!htbp]
  \centering
  \includegraphics[width=7.5cm]{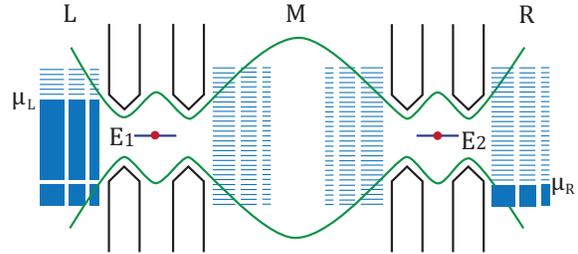}\\
  \caption{ Schematic setup for the transport through
  double dots which are connected
   by a central continuum. }\label{multi}
\end{figure}

Before solving the problem as shown schematically in Fig.\ 1,
we first elaborate the method to be used in this work.
As examined in Ref.\ \cite{Bran09}
for the simplest problem of single-dot transport,
either the usual Born-Markovian or non-Markovian master equation
is not good in general for finite-band reservoir/lead,
compared to the exact solution of this simplest model including
also the recently derived exact master equation approach \cite{ZhWM0810}.
In our present work, unlike using the Bloch-type rate equation derived
in Ref.\ \cite{SG98} under the conditions of WBL and large bias voltage,
we will apply an exact Schr\"odinger-equation-based
single electron wavefunction (SEWF) approach \cite{SG91,LG09}.
The SEWF approach is originated from an observation that, for noninteracting
system (transport), the many-particle state of the whole system
(i.e., the central system plus the two leads)
can be constructed by the Slater determinant
of all the single electron wavefunctions. Note that,
doing this, the Pauli principle is taken into account exactly.
In practice, the single electron wavefunction
is solved directly from the time-dependent Schr\"odinger equation.
Moreover, the initial single electron
states are summed over all the occupied states in the leads
which are defined by the Fermi surfaces.
In this way, the corresponding bias voltage is desirably introduced.
Knowing the wavefunction of the {\it entire} system, of course,
the current (or other quantities of interest)
can be calculated by the matrix element of the
corresponding operator and the total wave function.

While in wide band limit, it is indeed possible to recover
the master equation formalism from this SEWF
approach for {\it large} bias voltage\cite{SG91,LG09,Ora},
but unfortunately so far similar conversion does not succeed
for {\it small} voltage despite the perfect numerical agreement
with other exact approach (e.g., Ref.\ \cite{ZhWM0810}).
However, favorably, in practical manipulation
the SEWF approach allows us to calculate
the (transient) current directly from the single electron wavefunction,
not needing to construct firstly a master equation.
This is the most important advantage of the SEWF approach
applying to noninteracting transports.
As we will see in the following, in practice
the SEWF approach looks very like a generalization of the
scattering-matrix approach of the Landauer-Buttiker formulation.
That is, we first solve the single-electron state evolution
and calculate its current. Then, we apply
the well-known prescription to integrate the total current
between the two chemical potentials of the leads (the voltage window).
So in certain sense the present approach can be regarded
as a generalization of the Landauer-B\"uttiker formalism,
say, from the {\it stationary} to a {\it transient} version.

Now we turn to solving the DCD system shown in Fig.\ 1
by applying the SEWF approach.
For simplicity, we assume that each dot has only one
single spinless level.
For noninteracting dot, where the spin-up and spin-down
electrons transport independently,
this model simply corresponds to
a single level between the bias window.
In the presence of Coulomb interaction (inside the dot),
this model is also reasonable for the case of
a large Zeeman splitting so that within the (small)
bias window there is only one energy level with definite spin
(determined by the magnetic field orientation).
In this case no Coulomb correlation inside the dot
is relevant to the transport, since the dot level between
the bias window can be occupied at most by only one electron.
The whole setup is described by the following Hamiltonian
\beqn\label{H-2}
    &&H=\sum_l E_l\adg_la_l +E_1\adg_1 a_1 +\sum_m E_m \adg_m a_m \re
    &&+E_2 \adg_2 a_2+\sum_rE_r\adg_r a_r
    +\sum_l\Ome_l(\adg_1 a_l+\adg_l a_1)\re
    &&+\sum_m \big[ \Ome_m (\adg_1 a_m+\adg_m a_1)
    +\bar{\Ome}_m(\adg_2 a_m+\adg_m a_2)\big]  \re
    &&+\sum_r\Ome_r(\adg_2 a_r+\adg_r a_2) .
\eeqn
Here, $\adg_{1(2)}$ ($a_{1(2)}$), $\adg_{l(r)}$ ($a_{l(r)}$) and
$\adg_{m}$ ($a_{m}$) are, respectively, the creation (annihilation)
operators of the left (right) dot, left (right) lead
and the central continuum.
The first four terms are the free Hamiltonians of the individual parts,
while the last four terms describe their couplings,
i.e., between the dots and the leads and the central continuum,
with the respective coupling amplitudes of $\Ome_{l,r,m}$ and $\bar{\Ome}_{m}$.


In this work, following Ref.\ \cite{SG91}, we apply the SEWF approach
to study the transient transport behavior.
Let the initial state of the system correspond to filling the left
and right reservoirs at zero temperature with electrons up to the
Fermi energies $\mu_L$ and $\mu_R$, respectively.
For noninteracting transport, the problem can be solved exactly
for any values of the bias voltage, $\mu_L-\mu_R$.
Indeed, the total wave function for the noninteracting electrons,
$|\Psi(t)\rangle=\exp (-iHt)|\Psi(0)\rangle$, can be written at all
times as a product of single-electron wave functions,
$|\Psi(t)\rangle=\prod_{\bar{l}} |\psi_{\bar{l}}(t)\rangle$.
Note that, as well known, this product state in ``second quantization"
form is exactly equivalent to the construction of a Slater determinant.
Here $|\psi_{\bar{l}}(t)\rangle$ describes a single electron
initially occupying the state $\adg_{\bar{l}}\ket{0}$
in the left or right lead, with energy $E_{\bar{l}}\equiv E_{in}$.
We can express in general its state at time $t$ as a superposition
of all possible occupations in the whole system as follows
\beqn\label{WF-2}
    \ket{\psi_{\bar{l}}(t)}&=&\sum_lb_l(t)\adg_l\ket{0}+b_1(t)
    \adg_1\ket{0}+\sum_m b_m(t)\adg_m\ket{0}\re
    &&+b_2(t)\adg_2\ket{0}+\sum_r b_r(t)\adg_r\ket{0}  ,
\eeqn
where $\ket{0}$ denotes the {\it vacuum} without any electron,
and $\adg_j\ket{0}$ ($j=1,2,l,r,m$) are the single-electron
states with occupation amplitudes $b_j(t)$. Obviously, the
assumed initial condition is characterized by $b_l(0)=\del_{l\bar{l}}$,
while other amplitudes at the beginning are zero.
Note that all the amplitudes in \Eq{WF-2} depend on $\bar{l}$.
However, for brevity, we did not express this
dependence explicitly.

For the single electron initially in state
$\adg_{\bar{l}}\ket{0}$, the probability of finding it
in one of the leads or in the central continuum is therefore
$P_{L(R,M),\bar{l}}(t)=\sum_{l(r,m)}|b_{l(r,m)}(t)|^2$,
and the average current reads
$I_{L(R,M),\bar{l}}(t)=e\partial_t [P_{L(R,M),\bar{l}}(t)]$.
Consider $\mu_L>\mu_R$. At zero temperature,
the total current in each lead or into the continuum
can be expressed accordingly as
\begin{align}\label{I-LRM}
I_{L(R,M)}(t)=\int_{\mu_R}^{\mu_L} dE_{\bar{l}} \rho_L
      I_{L(R,M),\bar{l}}(t) \, ,
\end{align}
where $\rho_L$ is the density of states of the left lead.
More explicit result for the single electron current
$I_{L(R,M),\bar{l}}(t)$ will be presented
in the following respective examples.


Inserting the wavefunction, \Eq{WF-2}, into the time-dependent
Schr\"{o}dinger equation, we first obtain a set of coupled
linear differential equations for the amplitudes.
Then, applying the Laplace transformation,
$\tilde{b}_{j}(\ome)=\int^\iny_0 b_{j}(t)e^{i \ome t} dt$,
we convert the differential equations into the
following algebraic equations for $\tilde{b}_j(\ome)$:
\bsub\label{Lap2}
\beqn
    \ome\tilde{b}_l(\ome)-i b_l(0)=E_l \tilde{b}_l(\ome)
    +\Ome_l \tilde{b}_1(\ome)\\[1em]
    \ome\tilde{b}_1(\ome)=\sum_l\Ome_l \tilde{b}_l(\ome)
    +E_1\tilde{b}_1(\ome)+\sum_m \Ome_m\tilde{b}_m(\ome)\\
    \ome\tilde{b}_m(\ome)=E_m\tilde{b}_m(\ome)+\Ome_m
    \tilde{b}_1(\ome)+\bar{\Ome}_m\tilde{b}_2(\ome)\\[1em]
    \ome\tilde{b}_2(\ome)=\sum_m\bar{\Ome}_m
    \tilde{b}_m(\ome)+E_2\tilde{b}_2(\ome)+\sum_r\Ome_r\tilde{b}_r(\ome)\\
    \ome\tilde{b}_r(\ome)=E_r\tilde{b}_r(\ome)+\Ome_r\tilde{b}_2(\ome) .
\eeqn
\esub
This set of equations can be solved as follows.
First, find the solution of
$\tilde{b}_{l}(\ome)$, $\tilde{b}_{r}(\ome)$ and $\tilde{b}_{m}(\ome)$
in terms of $\tilde{b}_{1}(\ome)$ and $\tilde{b}_{2}(\ome)$,
from the first, third and fifth equations.
Then, substitute them into the second and forth equations.
In doing this, we will encounter the summation of
$\sum_{l(r,m)} |\Ome_{l(r,m)}|^2/[\ome-E_{l(r,m)}]$,
which can be further replaced by an integration
by introducing the density of states, $\rho_{l(r,m)}$,
for the leads and the central continuum.

In the case of wide-band limit (WBL), we can regard
$\rho_{l(r,m)}$ and $\Ome_{l(r,m)}$ as constants,
accordingly denoted as $\rho_{L(R,M)}$ and $\Ome_{L(R,M)}$.
We then have
\beqn\label{c-gam}
    \sum_{l(r,m)} \frac{|\Ome_{l(r,m)}|^2}{\ome-E_{l(r,m)}}
    =-i\frac{\Gam_{L(R,M)}}{2},
\eeqn
where the constant coupling rates,
$\Gam_{L(R,M)}=2\pi\rho_{L(R,M)}|\Ome_{L(R,M)}|^2$, are introduced.

In this work, of our greater interest is the more general case,
say, finite-band Lorentzian spectrum (FBLS),
for both the leads and the central continuum.
This can be modeled as
\beq\label{lapfactor}
    \Ome_{l(r,m)}=\Ome^{(0)}_{L(R,M)}
    \sqrt{\frac{\Lam_{L(R,M)}^2}{E_{l(r,m)}^2+\Lam_{L(R,M)}^2}} ,
\eeq
where the constant $\Ome^{(0)}_{L(R,M)}$ and $\Lam_{L(R,M)}$
characterize, respectively, the spectral height and width.
For the case $\Lam_{L(R,M)}$ much larger than the relevant energies,
$\Lam_{L(R,M)} \gg \omega$, the WBL is recovered. Otherwise,
the coupling rates in \Eq{c-gam} are of energy dependence as follows
\beq\label{L-w}
\Gam_{L(R,M)}(\omega)=\Gam^{(0)}_{L(R,M)}
\frac{i\Lam_{L(R,M)}}{\omega+i\Lam_{L(R,M)}} ,
\eeq
where $\Gam^{(0)}_{L(R,M)}=2\pi\rho_{L(R,M)}|\Ome^{(0)}_{L(R,M)}|^2$.

\vspace{0.3cm}
{\it Wide Band Limit}.---
In this case, from \Eq{Lap2}, we arrive at two equations for
$\tilde{b}_{1}(\ome)$ and $\tilde{b}_{2}(\ome)$:
\bsub\label{b12w}
\beqn
   &&\left(\ome-E_1+\frac{i\Gam_L}{2}+\frac{i\Gam_M}{2}\right)
   \tilde{b}_1(\ome)\re
   &&\hspace*{6em} =\frac{i\Ome_L}{\ome-E_{in}}-i\frac{\Gam_M}{2}
   \tilde{b}_2(\ome)\\
   &&\left(\ome-E_2+\frac{i\Gam_M}{2}+\frac{i\Gam_R}{2}\right)
   \tilde{b}_2(\ome)\re
   &&\hspace*{6em} =-i\frac{\Gam_M}{2}\tilde{b}_1(\ome) .
\eeqn
\esub
Here we assumed
$\Omega_M=\bar{\Omega}_M$ and $\Gamma_M=\bar{\Gamma}_M$.
While keeping $E_1=-E_2=\eps/2\neq 0$,
we assume further $\Gam_L=\Gam_R=\Gam_0$
and carry out an explicit analytic solution.
From \Eq{b12w}, we solve for
$\tilde{b}_{1}(\ome)$ and $\tilde{b}_{2}(\ome)$,
and find that commonly each has three poles,
$E_{in}$ and $\ome_{1(2)}=i(\pm\Del-\gam)/2$,
where $\Del=\sqrt{\Gam_M^2-\eps^2}$ and $\gam=\Gam_0+\Gam_M$.
Then, straightforwardly, an inverse Laplace transformation
leads to
\bsub\label{md-b}
\beqn
   && b_1(E_{in},t)
    =2\Ome_L\left[ f_{11}e^{-iE_{in}t}
    - f_{12} e^{-\frac{\gam}{2} t}  \right]  \re
    &&\\
    && b_2(E_{in},t)
    =2\Ome_L\Gam_M\left[-if_{21}e^{-iE_{in}t}
    + f_{22} e^{-\frac{\gam}{2} t}  \right] \re
    &&
\eeqn
\esub
In this solution, we introduced:
$f_{11}=(2E_{in}+\eps+i\gam)Z_1Z_2$,
$f_{12}=[(\Del-i\eps)Z_1 + (\Del+i\eps)Z_2]/(2\Delta)$,
$f_{21}=Z_1Z_2$,
and $f_{22}=(Z_1-Z_2)/(2\Delta)$,
where $Z_{1(2)}= e^{\pm\Del t/2} /[2E_{in}+i(\gam\mp\Del)]$.

With the knowledge of $b_{1(2)}(E_{in},t)$, one is able to carry
out the transient single-electron currents as follows.
As described previously above \Eq{I-LRM}, from the time derivative of
$P_{L(R,M),\bar{l}}\equiv\sum_{l(r,m)}|b_{l(r,m)}|^2$,
we can explicitly arrive to the following results:
$I_L(E_{in},t)=-e\Gam_L|b_1(E_{in},t)|^2-2e\Ome_L
\text{Im}[e^{iE_{in}t}b_1(E_{in},t)]$ for the left current;
$I_R(E_{in},t)=e\Gam_R|b_2(E_{in},t)|^2$ for the right current;
and $I_M(E_{in},t)= e\Gam_M|b_1(E_{in},t)+b_2(E_{in},t)|^2$
for the current flowing into the central continuum, from the both dots.
Here we would like to stress that in steady state these currents
satisfy the conservative relation $I_L=I_R+I_M$,
since the occupation probabilities in the two dots are
stationary (unchanged) in the steady state.
This conservative relation can be checked
numerically (in Figs.\ 2 and 3) or proven analytically.
The interesting feature is that, as we will see in the following,
$I_L\neq I_R$ even in the steady state!
This is somehow a novel consequence of the continuum spectral property,
but does not mean a breakdown of the electron number conservation.

\begin{figure}[!htbp]
  \centering
  \includegraphics[width=6cm]{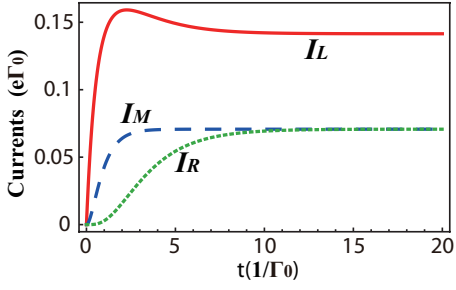}\\
  \caption{ Currents from an electron with incident energy
  in alignment with the dot level ($E_{in}=E_1=E_2=0$),
  under wide-band limit for both the leads and the central continuum. }
  \label{md-cur-1}
\end{figure}

While some (complicated) analytic results are available,
in Fig.\ 2 we numerically display the transient
single electron currents under the wide-band limit,
representatively for an incident energy $E_{in}$
in alignment with the dot level.
For simplicity, here and in the following numerical results,
we consider two identical dots and symmetric
coupling to the leads and the central continuum,
with $E_1=E_2$ and
$\Gamma_L=\Gamma_R=\Gamma_M=\bar{\Gamma}_M=\Gamma_0$.

\vspace{0.3cm}
{\it Finite-Band Lorentzian Spectrum}.---
We consider applying the finite-band Lorentzian spectrum (FBLS),
\Eqs{lapfactor} and (\ref{L-w}), to both the leads and the central continuum.
Again, for simplicity, we assume $\Gam^{(0)}_L=\Gam^{(0)}_R=\Gam_0$,
$\Gam^{(0)}_M=\bar{\Gam}^{(0)}_M=\Gam_M$,
and $\Lam_L=\Lam_R=\Lam_M=\Lam$.
Straightforwardly, solving Eq.\ (\ref{Lap2}) for
$\tilde{b}_1(\ome)$ and $\tilde{b}_2(\ome)$ gives
\bsub
\beqn
    \tilde{b}_1(\ome)&=&\frac{2i\Ome_{in}[\eps(\ome+i\Lam)+2\bet](\ome+i\Lam)}
    {(\ome-E_{in})G(\omega)}\\
    \tilde{b}_2(\ome)&=&\frac{2i\Ome_{in}\Lam\Gam_M(\ome+i\Lam)}
    {(\ome-E_{in})G(\omega)}
\eeqn
\esub
Here we introduced:
$\Ome_{in}=\sqrt{\frac{\Gam_0}{2\pi\rho}}
\sqrt{\frac{\Lam^2}{E_{in}^2+\Lam^2}}$,
$\bet=\ome^2-\Lam(\Gam_0+\Gam_M)/2+i\ome\Lam$,
and
$G(\omega)=4\bet^2-\Lam^2\Gam^2_M-\eps^2(\ome+i\Lam)^2$.
For double dots with aligned levels, $E_1-E_2=\eps=0$,
we find that $\tilde{b}_1(\ome)$ and $\tilde{b}_2(\ome)$
have, commonly, five poles in the complex $\ome$-plane:
$E_{in}$, $(\pm \gam_1-i\Lam)/2$, and $(\pm \gam_2-i\Lam)/2$,
where $\gam_1=\sqrt{2\Gam_0\Lam-\Lam^2}$ and
$\gam_2=\sqrt{2\Gam_0\Lam+4\Gam_M\Lam-\Lam^2}$.
For the purpose to express the {\it transient} current
in a more compact form, we define in this context
two types of inverse Laplace transformation:
$b_j(t)=\int\frac{d\ome}{2\pi}\tilde{b}_j(\ome)e^{-i\ome t}$, and
$\check{b}_j(t)= \int\frac{d\ome}{2\pi}\frac{\tilde{b}_j(\ome)}
{\ome+i\Lam}e^{-i\ome t}$, with $j=1,2$.
Then, after a short algebraic manipulations, the transient currents
from a single electron with incident energy $E_{in}$ can be calculated by
\bsub\label{mdcur}
\beq
    I_R(E_{in},t)=\Gam_0\Lam\text{Im}\left[b_2(t)\check{b}^*_2(t)\right]
\eeq
\beq
    I_L(E_{in},t)=-\text{Im}\left[\Gam_0\Lam b_1(t)
    \check{b}^*_1(t)+2\Ome_{in}b_1(t)e^{iE_{in}t} \right]
\eeq
\beq
    I_M(E_{in},t)=\Gam_M\Lam\text{Im}\{[b_1(t)+b_2(t)]
    [\check{b}^*_1(t)+\check{b}^*_2(t)]\}
\eeq
\esub
Here we omitted the energy $E_{in}$ in $b_{1(2)}(E_{in},t)$
and $\check{b}^*_{1(2)}(E_{in},t)$, for the sake of brevity.
Integrating the results of \Eq{mdcur} over $E_{in}$ within a proper energy
window, transient currents for arbitrary bias voltage can be obtained.

Before turning to the detailed behaviors of the currents,
we first make a general remark on the transfer nature
through the {\it continuum}.
As we have mentioned in the introductory part, typically,
a quantum particle cannot reenter the dots
after it irreversibly decays into a continuum.
To make this statement more specific, we may consider
a single level (bound state) coupled to a continuum.
In this case, the continuum can be a Fermi reservoir,
but with the Fermi surface much lower than the bound-state level.
Then, it is a common knowledge that an electron,
initially occupying the bound state,
will irreversibly decay into the reservoir along time,
and will never come back in long-time limit.
By contrast, for the case of DCD setup, the electron
can transmit through the central continuum
even in long time limit \cite{SG11a}.
Below we explain that the basic reason is owing to
the {\it coherent coupling} of the two dots with the continuum.
In this case,
if we convert the above wavefunction approach/solution under
into a rate equation,
we will find that the (state) occupation transfer
is mediated by a {\it quantum coherence} term,
i.e., by the off-diagonal elements of the density matrix.
This means that, if the ``quantum coherence" is destroyed,
the electron cannot propagate from one dot to another through the continuum.
We demonstrate this feature, more transparently, as follows.
In wide band limit, we can construct a Lindblad-type master
equation for the coupling of the two dots to the continuum \cite{SG11b}:
$\dot{\rho} = -i[H_S,\rho] + \Gamma_M {\cal D}[a_1+\chi a_2]\rho $.
Here $\rho$ is the reduced density matrix of the double dots;
the Lindblad super-operator reads
${\cal D}[a] \rho=a\rho a^{\dg}-\frac{1}{2}\{a^{\dg}a,\rho\}$;
and $\chi=\bar{\Omega}_{M}/\Omega_{M}$ is introduced.
We conclude that the coherent superposition of $a_1$ and $a_2$
in the Lindblad super-operator is of crucial significance
to the charge transfer through the continuum,
which actually provides such cross terms as $a_1\rho a^{\dg}_{2}$
to mediate the passage through the continuum.
Otherwise, if we destroy the {\it coherence} by
removing this type of cross terms and keep only
terms as ${\cal D}[a_1] \rho$ and ${\cal D}[a_2] \rho$, we will find that
the electron cannot transfer between the dots through the continuum.

\begin{figure}[!htbp]
  \centering
  \includegraphics[width=7.8cm]{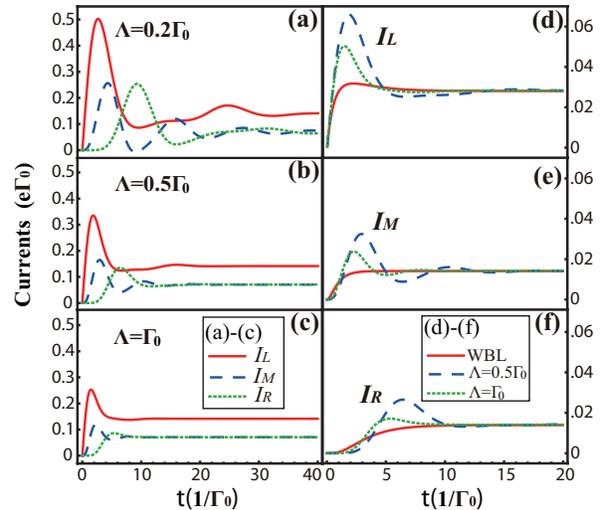}
  \caption{ (a)-(c):
  Single electron currents for finite-band Lorentzian spectrum, with bandwidths
  of $\Lam=0.2\Gam_0$, $0.5\Gam_0$ and $\Gam_0$, respectively.
  The electron's incident energy is aligned with the dot level
  which locates at the center of the Lorentzian.
  (d)-(f): Integrated currents over an energy window
  $(-0.1\Gam_0,0.1\Gam_0)$ around the dot level ($E_1=E_2=0$).  }
  \label{md-cur-2}
\end{figure}


In Fig.\ 3 we show the result from the FBLS which, qualitatively,
differs from the result under the WBL as shown in Fig.\ 2.
That is, the current under WBL does not display oscillations with time,
while it oscillates in the case of FBLS.
This oscillation behavior is indeed originated from
an essential non-Markovian effect.
The difference in Fig.\ 2 and Fig.\ 3(a)-(c) is
out of a simple intuition:
since the incident energy is in resonance with the dot level,
we may expect the electron's motion to be affected mostly by
the continuum states with {\it similar energy} of the incident electron.
However, by altering the bandwidth of the Lorentzian spectrum,
the result in Fig.\ 3(a)-(c) indicates that the electron's motion
is also influenced strongly by the ``remote" continuum states
with quite different energies.

The reason of having stronger non-Markovian behavior when coupled to
a narrower band continuum may be understood in certain sense by imagining
an extreme situation: if the narrow band {\it continuum} is reduced to
discrete levels in the same energy range,
{\it reversible} quantum coherent oscillations
will replace the oscillating decay behavior.
There exists, however, an essential difference between them.
That is, no matter how narrow the continuum is,
it will lead to an {\it irreversible} decay into the continuum.
Of particular interest is that the electron
is more easily coming back during the decay process,
for a narrower bandwidth of the continuum.
This understanding provides a {\it dynamic} insight for the nature
of the non-Markovianity, which should be equivalent to
the sophisticated {\it information-theoretic} notion by relating
the non-Markovianity with a reversed flow of {\it information}
from the environment back to the system of interest \cite{Br09,Cir08}.

We can further understand the stronger transient non-Markovian effect
for a narrower band of continuum by means of the time-energy uncertainty
relation. On relatively shorter timescale,
the electron couples to a larger range of energies of the continuum.
Then, for a wide band continuum, more ``remote" energy states will
involve into the (tunnel) coupling dynamics, resulting in a
destructive interference cancelation which destroys the ``coming back"
process (oscillating behavior).
While on longer timescale (in steady state), the result is less
sensitive to the bandwidth, as shown in Fig.\ 3(a)-(c).
However, we find that the steady-state result under the WBL can
be faster reached with a small increase of the bandwidth,
while the short-time behaviors remain to be quite different.
This feature demonstrates that the bandwidth effect
hides in a more pronounced manner in the transient dynamics.
In this context, we see that the {\it transient}
single electron current captures more information
than the conventional {\it stationary} scattering approach.
To make the result more reasonable,
we further integrate the current over a small energy window
around the dot level and display the result in Fig.\ 3(d)-(f),
where we find that the finite-bandwidth induced non-Markovian
behavior is still evident.

Finally we discuss the interesting
steady-state behavior of the DCD system.
The results in both Figs.\ 2 and 3 show that, even in steady state,
the left and right currents are {\it unequal}, i.e., $I_L > I_R$.
This feature distinguishes the central {\it continuum},
drastically, from any {\it finite-size} well, since in the latter case
the steady-state left and right currents {\it must} equal to one another.
The unequal left and right currents indicate a drastic transition
when the finite-size central well approaches to a {\it continuum} limit.
Actually, the essential physics is the development of {\it irreversibility}
in this limit, which causes part of the transport electrons
being irreversibly lost into the continuum and floating there,
while other electrons supporting the transport current
due to the {\it coherent-coupling-mediated transfer} between the dots.

In this context one may expect a formation of chemical potential
in the continuum, owing to electron filling up there.
Then, in steady state the net current into the continuum is zero.
Of interest is that,in the central continuum,
we did not introduce any {\it inelastic scattering},
which therefore differs from what happens
in the electrodes in transport, where
the inelastic scattering processes (quickly) relax
the electrode to a local equilibrium.
For the central continuum in our DCD system,
the absence of inelastic scattering implies that the electrons
stayed in the continuum just float there,
with no energy relaxation to form a local equilibrium
with well-defined Fermi energy.
We notice that, to account for phase breaking in quantum transport,
B\"uttiker once proposed a type of dephasing model
by introducing a virtual {\it side reservoir} \cite{But86,Li0201}.
In that approach, the transport electron first enters the side reservoir,
experiences inelastic scattering in it, then returns back into the {\it system}.
As a result, the quantum coherence of
the electron is destroyed or partially destroyed.
In B\"uttiker's dephasing model,
the side reservoir is assigned
an appropriate chemical potential to guarantee
a {\it zero} net current flowing into the side reservoir.
On the contrary, in the DCD system, the {\it nonzero} net current
flowing into the central continuum is
a novel consequence of the continuous spectral property.

\vspace{0.3cm}
{\it Discussion and Summary.}---
Before summarizing the work, we would like to present a discussion
on the non-Markovian nature of the SEWF approach.
Indeed, conventionally, the non-Markovian feature is manifested
in terms of a {\it time-nonlocal} master equation.
The time-nonlocal feature, as is well known, reflects a memory effect.
However, there is an alternative form of non-Markovian master equation,
which is actually exact for some solvable models \cite{ZhWM0810}.
In that form, the non-Markovianity is manifested by a time-dependent decay
rate, which depends on the entire past history thus keeps the full memory effect.
Concerning the SEWF approach, it is built on a set of infinite
coupled differential equations for the dot- and continuum-state
amplitudes, see \Eq{Lap2}.
The basic solving procedure, in time domain,
is integrating first the continuum amplitudes,
then substituting them into the equations for the dot-state amplitudes.
In doing this, the dot-state evolution depends on the
entire past history of the continuum/reservoir, thus holds a memory effect.
This feature, in essence, corresponds to the time-dependent decay rate
in the exact master equation formalism \cite{ZhWM0810}.

While the non-Markovian behavior is rather obvious
for the finite-bandwidth Lorentzian spectrum,
it is of interest to understand that,
in the case of WBL, the transport under {\it small}
bias voltage is in fact also non-Markovian,
despite that the underlying non-Markovianity is usually quite weak,
as indicated by $I_L$ in Fig.\ 2.
Notice that, the conventional master equation is obtained
by summing over all the single electron states
initially occupied in the leads,
which are actually defined by the Fermi surfaces.
While for each individual electron, starting its evolution
from a state in the lead, the transient dynamics is non-Markvian
in essence as analyzed above,
whether or not the master equation (obtained after the summation of
the single electron states) is Markovian in WBL depends on the bias voltage.
That is, under large bias voltage it is Markovian \cite{SG98}
while under small voltage, using the method in Ref.\ \cite{ZhWM0810},
it can be shown that the corresponding master equation
is non-Markovian even in WBL.
This latter result is in agreement with the non-Markovian nature
of the {\it exact} SEWF approach.

To summarize, in this work we have presented an exact solution for the
non-Markovian transport through a double-dot connected by a continuum.
Unlike the intuition that a quantum particle does not reenter
the dots after its irreversible decay into the continuum,
we demonstrate a highly non-Markovian quantum-coherence-mediated
transmission through the continuum, owing to coherent coupling
of the two dots with the same continuum.
The continuous spectral nature of the continuum also supports a steady
state with unequal source and drain currents, as a result of electrons
irreversibly entering the continuum and floating in it.

\vspace{0.2cm}
{\it Acknowledgments.}---
We thank S. Gurvitz for stimulating discussions.
This work was supported by the NNSF of China
(No. 101202101 \& 10874176),
and the Major State Basic Research Project of China
(No.\ 2011CB808502 \& 2012CB932704).


\end{CJK*}

\begin{references}
\bibitem{note-1}
Note that, under small bias voltage, the transport is in essence
non-Markovian even in the case of wide-band limit.
\end{references}
\end{document}